\documentclass[10pt,aps,prd,twocolumn,preprintnumbers,superscriptaddress,nofootinbib,floatfix]{revtex4-2}
\usepackage[utf8]{inputenc}
\usepackage{amssymb}
\usepackage{tensor}
\usepackage{xspace} 
\usepackage{graphicx}
\usepackage[caption=false]{subfig}
\usepackage{breakcites}
\usepackage{ragged2e}
\usepackage{amsmath}
\usepackage{multirow}
\usepackage{academicons}
\usepackage{orcidlink}

\usepackage{setspace}

\usepackage{hyperref}
 
\hypersetup{
colorlinks,linkcolor=blue,citecolor=blue,urlcolor=blue
}

\usepackage{natbib}
\usepackage{tablefootnote}
\usepackage{footnote}
\usepackage{amsmath}
\usepackage{amssymb}
\usepackage[T1]{fontenc}

\begin{document}

\title{Towards an exact approach to pulsar timing}

\newcommand{\INAF}{\affiliation{INAF, Osservatorio Astronomico di Cagliari, Via della Scienza 5, 09047 Selargius (CA), Italy}}

\author{Amodio Carleo \orcidlink{0000-0001-9929-2370}}
\email{amodio.carleo@inaf.it}
\INAF

\author{Delphine Perrodin}
\INAF

\author{Andrea Possenti}
\INAF

\date{\today}


\begin{abstract}
 The pulsar timing technique, which compares the observed arrival times of electromagnetic radiation from a pulsar with the predicted arrival times derived from a theoretical model of the pulsar system, is used in pulsar astronomy to infer a multitude of physical information and to constrain possible corrections to General Relativity (GR). The propagation delay is usually computed using formulas based on a post-Newtonian approach, for both the light trajectory and the orbital motion. However, evidence has recently emerged that this approximation may no longer be sufficient when the companion object is a supermassive black hole; deviations from a full GR computation of the propagation delay can reach a few seconds. In this paper, we analyze the case of binary pulsars with a stellar or intermediate black hole companion, whose discovery and timing are key goals of SKA. With a numerical algorithm, we have found that in this case, the full GR value depends only on the semi-major axis of the relative orbit and on the mass of the black hole companion. If the mass of the latter is sufficiently large ($100 M_{\odot}$), the maximum difference between the two approaches is significant ($\sim10^{-7}$ s) even for large binaries  ($\sim10^{16}$ cm), and increases up to $\sim 10^{-4}$ s when  the mass is $10^5 M_{\odot}$. We also consider relativistic corrections to the orbital motion, and discover that they can strongly affect the value of the propagation delay. We conclude that in the future, post-Newtonian formulas should be replaced with a more accurate approach in these systems, especially in view of future discoveries made by new large telescopes such as SKA. 
\end{abstract}



\maketitle

\section{Introduction}
Neutron stars are compact objects that are formed following the explosion of massive stars. Radio pulsars are neutron stars that emit beams of radio waves, which are detected at Earth's radio telescopes as extremely regular series of `pulses', corresponding to the extremely regular rotation period of pulsars. The emitted energy, although predominant in the radio band, is actually broadband, with energies up to the GeV range and sometimes even in the TeV range. In some cases, pulsations appear in different energy bands, even if peaks in different bands may be associated with different emitting regions \cite{Spolon:2018qyf}. Since they are remarkably precise clocks, pulsars can be used to investigate many different aspects of physics, from the interior of neutron stars (NS) \citep{Ascenzi:2024wws}, to GR tests \citep{Stairs_2003,Freire_2024}, the density of the interstellar medium \cite{Keith:2024kpo}, dark matter \cite{Smarra:2024kvv} and the gravitational wave background (GWB) \cite{Verbiest:2024nid}. Most of the aforementioned studies involve the so-called “pulsar timing” technique, i.e. the measurement of the times-of-arrival (ToAs) of the radio signals emitted by the pulsar, which are then compared to the TOAs predicted by a  theoretical model that accounts for many astrophysical processes involving the pulse's emission, among which: the pulsar
system dynamics (e.g. pulsar spin, proper motion, etc.),
solar system dynamics (e.g. motion of the Earth and planets), and the effects of the interstellar medium (e.g. dispersion and scintillation) or solar wind effects. Depending on whether the pulsar is isolated or in a binary, this multi-parameter fit gives several important parameters (the so-called ephemeris), such as the spin period, spin period derivative, orbital period (if in a binary), position in the sky, eccentricity, etc. Currently, the pulsar community uses three main data analysis packages to analyze ToAs, namely \texttt{TEMPO} \cite{TEMPO}, \texttt{TEMPO2} \cite{Hobbs:2006cd} and \texttt{PINT} \cite{PINT}. Since the latter was written completely independently, it offers an excellent tool for cross-checking the output. This is becoming necessary, since for high-impact precision timing programs, such as gravitational wave detection efforts, it is critical to compare results from more than one single data analysis pipeline. \\
The first hint of the power of this method was seen in the case of the binary system PSR B1913+16, whose orbital decay agrees with the predicted GR values to better than 0.5\% \cite{taylor1989New}: the observed accumulated shift of periastron is in excellent agreement with GR. Over the years, there have been numerous studies on possible violations of GR using the timing of pulsars  \citep[e.g.][]{Stairs_2003,https://doi.org/10.48550/arxiv.2204.13468}, revealing that pulsars are a great research tool in this field. As a result, many alternative gravity theories have been strongly constrained or even falsified (see \cite{Freire_2024} for a recent review). Pulsars also provide a way to test the no-hair theorem as well as the
cosmic censorship conjecture \citep{Liu_2012,Izmailov:2019cqr}. More recently, pulsar timing has been exploited for the detection of the low-frequency gravitational wave universe \cite{InternationalPulsarTimingArray:2023mzf} through the combination of the timing data for an array of approximately 100 millisecond pulsars (MSP) observed by the largest radio telescopes in the world  \cite{Verbiest:2024nid}. Gravitational waves cause changes in the travel times of radio signals between pulsars and the Earth, and a Gravitational Wave Background (GWB) is predicted to produce a specific pattern (the Hellings \& Downs curve) in the correlated timing residuals of pairs of pulsars. Generally, any unmodeled effect will appear in the timing residuals: trends in the residuals are indicative of non-optimized parameters, while white noise residuals with low root-mean-square suggest a good timing model.  In the case of a GWB, we expect red noise in the timing residuals at nHz frequencies \citep{detweiler}. In 2023, evidence for a possible GWB was revealed, and looks very promising \citep{NANOGrav:2023gor, EPTA2023, PPTA2023}. \\
 
A widely adopted timing model that includes all relativistic effects in the dynamics of a binary system up to the first post-Newtonian level is
based on Damour and Deruelle’s approach  \citep{deruelle} (DD), which has allowed the scientific community to perform self-consistent tests of gravity by means of the parametrised post-Keplerian (PPK) formalism. The main feature of this approach is the closed-form solution of the relativistic first post-Newtonian (PN) two-body problem, making its application immediate. The same approach can be used to incorporate other effects, such as the spin-induced quadrupole moment of the companion \cite{Wex:1998} and tidal deformations \cite{voisin}, as well as a combination of all these effects. A pseudo-DD formalism can also be adopted, in specific conditions, to include a generic perturbation to Keplerian motion of the form $\Phi(r)_{\alpha} \propto -J r^{-\alpha}$, where $J \gg 1$ is the dimensionless perturbation parameter, $r$ the radial coordinate and $\alpha$ the perturbation index \cite{voisin}. The DD model is therefore the ensemble of pulsar timing formula with the DD description for the binary orbit. In addition to the description of the orbit, which can be Keplerian or post-Keplerian, one usually adopts a post-Newtonian description for the space-time of the binary, in order to account for relativistic effects of the emission time and travel time of the pulsars' emitted photons. In the weak-field limit, there are different time delays. The three most important ones are: the Roemer delay, the Shapiro delay and the Einstein delay, which need to be taken into account both in the Solar System and in the pulsar binary. In the pulsar binary, the Roemer delay is given by the geometric variation of the distance between the emission point and the observer due to the orbit of the pulsar around the companion. In the Solar System, a similar and additional effect is due to the Earth's motion around the Sun; in order to avoid the modulation induced by Earth's orbit, ToAs are referred to the Solar System barycenter (SSB), where coordinate time is defined as $t_{SSB} = t_{em} + (1/c)|\mathbf{r}_{p} -\mathbf{r}_{b}|$, where $t_{em}$ is the time of photon emission, and $\mathbf{r}_{p,b}$ is the position of the pulsar or SSB (usually calculated using distant quasars). The Shapiro delay is the (always positive) additional delay that takes into account the deviations of light caused by the gravitational field of the Sun (in the Solar System) and the pulsar's companion (in the binary); it is easily obtained in the weak-field approximation by solving the geodetic equations and is known up at to the 2PN order. Corrections to include lensing and geometric effects are also known \cite{Rafikov2005}. Finally, the Einstein delay is due to the difference between proper time and coordinate time at Earth (in the Solar System) and at the pulsar (in the binary), which is different because of the influence of the companion's gravity as well as time dilation. Sometimes, the sum of Roemer and  Shapiro delays is called the `propagation delay'. These post-Newtonian formulas are widely used by the pulsar community, even if binary pulsars are  strong-gravity systems. The reason is that the accuracy of radio data, quantified by the root-mean-square (RMS) value of the timing residuals, although good, is usually not good enough to reveal second-order effects. In exceptional cases, such as in the Double Pulsar system, higher-order corrections to periastron advance and Shapiro delay formulas are necessary \cite{Kramer_2021}.   \\

Nowadays, the timing precision of some pulsar experiments has increased from $\sim10$ $\mu$s to  $\sim100$ ns \cite{Wang_2024}, with good reasons to expect residuals below $\sim50$ ns in the coming years \cite{Hu_2022}, even near superior conjunction, where accuracy is highly compromised due to eclipses. An accuracy of $\sim100$ ns or better is also a minimum requirement for detecting a stochastic GWB \cite{Stappers_2018}. Even if software strategies play an important role in precision improvement \cite{Wang_2024}, the real leap in quality will come with the entry into operation of SKA, which could achieve an accuracy of $\sim10$ ns \cite{Smits_2011,Liu_2011}, allowing for extraordinarily accurate measurements of the distance to numerous pulsars, as well as the detection of single sources of nanohertz gravitational waves. One of the goals of SKA is also to detect and perform the timing of pulsars in the galactic center (GC), which is a discovery that has been awaited for many years and which has numerous scientific implications: in addition to an independent measurement of Sgr A*'s mass, pulsar timing analyses in the GC would allow for the measurement of its spin magnitude \cite{Zhang_2017}, quadrupole moment \cite{Liu2012}, the structure of dark matter \cite{DM:2023ubk}, as well as a series of tests on the geometry of space-time (like the validity of the no-hair theorem \cite{nohairTh_2019}) and corresponding gravity tests, given the enormous amplification of any deviations from Einstein's theory \cite{Liu_2012, DellaMonica:2023ydm}. However, pulsar searches with SKA are not limited to the GC; another main goal is to find, and time, with a similar precision, a pulsar in orbit around a \textit{stellar} black hole \cite{Desvignes_2016}, which is the type of binary system this paper will focus on.\\
Given the strong-field regime in the GC region, one wonders whether the usual post-Newtonian timing formulas are valid or not (as compared to the accuracy of the data). All of this forces us to re-evaluate the theoretical framework based, as mentioned above, on a post-Newtonian approximation both for the orbital dynamics and to model the spacetime of the companion object. The question can then be extended to binary systems that are not necessarily extreme (i.e. a supermassive black hole (SMBH) + pulsar), but instead composed of a pulsar and a stellar black hole (BH). The answer to this question is the main goal of this work. \\

The first evidence for a discrepancy between the full GR treatment and the post-Newtonian approximation in pulsar timing was covered in \cite{Eva_2019} and then in \cite{Kimpson2019,Kimpson2020}. In particular, in \cite{Eva_2019}  an
exact analytical formula for the propagation delay in a Schwarzschild space-time is shown. According to this paper, for (nearly) edge-on orbits, the Shapiro delay with a lensing correction deviates quite significantly from the actual, exact propagation delay. For Sgr A*, the maximum difference (superior conjunction) between the exact (analytical) formula and the sum of the single post-Newtonian delays is of $\sim1$ s for a pulsar orbiting at distance $r=10^2-10^3$ M, where M is the mass of Sgr A*. When the orbit is inclined, the difference decreases quickly, and arrives at $\sim10^{-1}$ s for an orbit with the same radius but with an inclination of $\pi/3$. The issue was then further explored in \citep{Bilel}, where a comparison between an exact analytical result for the frame-dragging  delay (an additional quantity to the propagation delay) and two post-Newtonian derivations for this effect was analyzed. The conclusion is that post-Newtonian based treatments overestimate the frame dragging effect on the light-like signals, in particular around and after superior conjunction, hence the analytical solution provides a more reliable and accurate approach: if the dimensionless spin of Sgr A* is $a=0.9$, then the best post-Newtonian formula for the frame-dragging delay \cite{Rafikov2005} for a  pulsar at $r=10^3$ M from the SMBH predicts a delay which is 3 s larger than the full GR computation. However, this paper does not report estimates of the total propagation delay, focusing only on the effect of the spin. Later, in \cite{Carleo:2023qxu}
 the effect of `matter’ on the propagation time was investigated using a similar approach, showing that the presence of dark energy around Sgr A* would cause an additional delay of  $\simeq300$ s at superior conjunction (with an advance of $\simeq40$ s at the inferior conjunction) for  SGR J1745-2900, the closest magnetar orbiting Sgr A*. However, at the moment the timing of SGR J1745-2900 has not yet reached sufficient levels
of precision for this purpose, mainly due to the intrinsic
variability of the source. These papers use an analytical approach to compute the propagation delay, but the necessity of a fully relativistic treatment has  also emerged in numerical analyses \cite{Zhang_2017,Kimpson2020}. A  common feature in these works is the assumption of a pulsar orbiting a SMBH, therefore allowing the use of the one-body approximation, since the neutron star is considered as a test particle in the space-time of the black hole. This circumstance, however, in addition to being unique (there is only one supermassive black hole in our galaxy) still does not have an observational counterpart, while the number of pulsars orbiting stellar black holes ($5-100$ $M_{\odot}$) should be relatively higher, especially in globular clusters \cite{Kremer_2018}. The first discovery of this type may be PSR J$0514-4002$E, an eccentric binary millisecond pulsar in the globular cluster NGC 1851 with
a total binary mass of $3.887 \pm 0.004$ $M_{\odot}$, i.e. with a companion in the mass gap between a very-massive NS and a low-mass BH \cite{Ridolfi_2022,Barr_2024}. In addition, in globular clusters, we also expect intermediate-mass black holes ($10^2-10^5$ $M_{\odot}$) \cite{Fujii_2024}, for which post-Newtonian approximations could be not accurate enough.     \\

Here we present the results of an analysis on the discrepancy in the propagation delay (the most substantial part of TOAs) between the usual post-Newtonian formula and a full GR approach, having in mind close-to-discovery systems composed of a pulsar and a stellar or intermediate black hole. The outline of the paper is as follows: in Section 2, we recover the geodesic equations from which the propagation delay is computed; in Section 3, we briefly recall the post-Newtonian formulas; in Section 4, we illustrate our computational algorithm for the full GR computation, with the relativistic case with corrections to the orbital motion discussed in Section 5.  We show our results in Section 6 and  close the manuscript with a summary and outlook in Section 7.  \\

\section{The propagation delay in general relativity}

In this section, we refer to the results in \cite{Eva_2019,Bilel,Carleo:2023qxu} in order to compute the propagation delay in Kerr space-time. This is the time it takes for the radio pulses emitted by a pulsar to reach an observer due to various factors affecting the propagation of the signal through space. This delay is a critical component in precisely measuring the arrival times of pulses and accurately modeling the pulsar's behavior. In addition to the aforementioned Roemer and Shapiro delays, the propagation of the signal is generally affected by several effects, such as the interaction with free electrons of the interstellar medium (causing the lower-frequency components of the signal to slow down more than the higher-frequency components), with Earth's ionosphere and the solar wind. Since these are usually smaller effects and, ignoring the transformation from coordinated time to proper time, one usually speaks of `propagation delay' referring mainly to the sum of Roemer and Shapiro delays. In a full GR treatment, such a distinction  is not possible and one only solves the photon geodesic equations, obtaining the amount of (coordinate) time required for the pulse to propagate from the NS to the observer in spacetime. About the latter, it is common to consider only the companion object as the source of the stress-energy tensor $T_{\mu\nu}$ to delineate the metric, even if in general both the objects contribute to $T_{\mu\nu}$. In GR, an exact metric for a binary system does not exist due to the non-linearity of the theory and only  approximate or numerical solutions can be found. However, if the companion is a black hole, it is sufficient to ignore the mass of the pulsar in the computation of the photon delay. The case of the trajectory of the pulsar itself is different: if the pulsar's mass is negligible with respect to the companion's mass, then it moves on time-like geodesics of the companion space-time; if masses are comparable or not so different, then the motion must be corrected by considering at least 1PN terms. \\
In order to compute the propagation delay, we focus on the companion object and describe the space-time with a Kerr metric, which is given in Boyer-Lindquist coordinates ($t,r,\theta,\varphi$) by ($G=c=1$)
\begin{equation}\label{eq1}
\begin{array}{ccc}
d s^{2}=-\Big(1  -\frac{2 M r}{\Sigma^{2}}\Big) d t^{2}  +\frac{\Sigma^{2}}{\Delta} d r^{2} \\[6pt] -\frac{2 a \sin^{2} \theta\left(2 M r\right)}{\Sigma^{2}} d \varphi d t 
+\Sigma^{2} d \theta^{2} \\[6pt]
    +\sin^{2} \theta\left(r^{2}+a^{2}+a^{2} \sin^{2} \theta \frac{2 M r}{\Sigma^{2}}\right) d \varphi^{2}
\end{array}
\end{equation}
where we defined
\begin{equation*}
 \Delta\doteq r^{2} -2Mr +a^{2}, \; \; \;  \Sigma^{2}\doteq r^{2}+a^{2}\cos^2{\theta}, \end{equation*}
and $M$ is linked to the mass $\bar{M}$ of the object (likely a black hole) by $\bar{M}=(c^2/G) M$ while $a=J/(Mc)$ where $J$ is its angular momentum.  Using the metric components $g_{\mu\nu}$ from the line element $ds$,  geodesic equations are given by 
\begin{equation}\label{eq2}
    g_{\mu \nu} \dfrac{dx^{\mu}}{d\tau}\dfrac{dx^{\nu}}{d\tau}=\epsilon
\end{equation}
where $\tau$ is an affine parameter and $\epsilon=0,-1$. While in Schawrzschild space-time orbits starting in the plane (for example $\theta=\pi/2$) remain planar, in Kerr metric this is not true and an additional motion constant is needed: in addition to energy E and angular momentum M, one also needs a separation constant called Carter's constant. Using the Hamilton-Jabobi formalism, we obtain the well-known   geodesic equations in Kerr space-time:
\begin{equation}\label{eq:t}
    \dot{t} = \dfrac{(r^2+a^2)(r^2+a^2-a\lambda)}{\Delta}-a(a-\lambda)+a^2\cos^2{\theta}
\end{equation}
\begin{equation}\label{eq:varphi}
   \dot{\varphi} = \dfrac{a(r^2+a^2-a\lambda)}{\Delta}-a+\dfrac{\lambda}{\sin^2{\theta}}
\end{equation}
\begin{equation}\label{eq:teta}
   \dot{\theta}^2 = q + \cos^2{\theta} \left[ \left(1 + \dfrac{k}{E^2}\right)a^2 -\dfrac{\lambda^{2}}{\sin^2{\theta}} \right] =  \Theta(\theta)  
\end{equation}
\begin{equation}\label{eq:r}
     \dot{r}^2 = -\Delta\left[q-\dfrac{k}{E^2}r^2+(\lambda -a)^2 \right] + \left( r^2+a^2-\lambda a \right)^2 = R(r) .
\end{equation}
Here, a dot means derivative w.r.t. the so called Mino time $\gamma$, which satisfies the condition $dx^{\mu} = (\Sigma/E) p^{\mu} d\gamma$, while $q\doteq  \mathcal{C}/E^{2}$. Notice that quantities $m$, $E$, $L$, $\mathcal{C}$ are  constants of motion for the EOMs (\ref{eq:t})-(\ref{eq:r}). For the simplest case of equatorial photon orbits ($q=0$), the radial potential $R(r)$ has four real roots, namely $r1<r2<r3<r4$, or at most 2 complex roots,  depending on whether the impact parameter $\lambda$ is greater or smaller than a certain critical value $\lambda_c$ (one for co-rotating and one for counter-rotating orbits for  Kerr metric) corresponding to the so-called unstable spherical orbit.  The above equations can be analytically solved using elliptic integrals in Legendre form. In particular, combining Eqs. (\ref{eq:t}), (\ref{eq:teta}) and (\ref{eq:r}), one gets 
\begin{equation}\label{eq20}
    t_{a}-t_{e} = \int_{\gamma_{r}} \dfrac{G(r)}{\Delta \sqrt{R(r)}} dr + \int_{\gamma_{\theta}} \dfrac{a^2 \cos^2{\theta}}{\sqrt{\Theta (\theta)}} d\theta
\end{equation}
where we have defined
\begin{equation}\label{eq21}
    G(r) = r^2 (r^2+a^2+ac(a-\lambda)) + 2Mra(a-\lambda).
\end{equation}
The integral path $\gamma_{r}$ starts at the radial point of emission $r_e$ and either runs monotonically increasing to infinity\footnote{We assume an observer at infinity at $\varphi_a=0$.} (direct trajectory) or first decreases in radius towards a turning point outside of the horizons at $r=r_4$ and then continues towards infinity (indirect trajectory). Therefore, the radial integral is split as 
\begin{equation} \label{eq22}
    \left( \int_{r_4}^{\infty}  \pm \int_{r_4}^{r_e}\right) \dfrac{G(r)}{\Delta \sqrt{R(r)}} dr
\end{equation}
where the minus sign is for a direct trajectory and the plus sign for a flyby motion. Note that when $\lambda<\lambda_c$ only a direct path is possible, while when $\lambda>\lambda_c$ both paths are possible according to the emission position. Neglecting the latitudinal motion (equatorial orbits) and using elliptic integrals, we obtain
\begin{equation}\label{eq24}
\arraycolsep=1.4pt\def\arraystretch{2.2}
\begin{array}{ccc}
t_{a}-t_{e} =&  T_{r}(\infty,\lambda_e) \pm T_{r} (r_e,\lambda_e) \\
\end{array}
\end{equation}
where $\lambda_e$ is the impact parameter of the emission point and we defined \cite{Eva_2019,Bilel,Carleo:2023qxu} 
\begin{equation}\label{eq-26}
\arraycolsep=1.8pt\def\arraystretch{2.6}
\begin{array}{l}
T_{r}(r, \lambda)=\delta  \Big[F(x, k)  \Big(4 M^2   \lambda \\
+2 M r_3  
+\frac{1}{2}  \big[r_1\left(r_3-r_4\right)+r_3\left(r_3+r_4\right)\big]+\frac{B_+ l}{l_+}+\frac{B_- l}{l_-}\Big) \\
-\frac{1}{2} E\big(x, k\big) \big(r_4-r_2\big)\big(r_3-r_1\big) 
+\Pi(x, l, k) \Big(2 M r_4 -2 M r_3 \Big) \\
+\Pi(x,l_{+}, k) \Big(B_{+}-\frac{l B_{+}}{l_{+}}\Big)+\Pi\big(x, l_{-}, k\big) \Big(B_{-}-\frac{l B_{-}}{l_{-}}\Big)\Big] \\
+\frac{\sqrt{R(r)}}{r-r_3}  \\
\end{array}
\end{equation}
where $r_{\pm}$ are the horizons, $ \delta = 2/\sqrt{\left(r_4-r_2\right)\left(r_3-r_1\right)}$, $l = (r_1-r_4)/(r_1-r_3)$, $l_{\pm} = l (r_3-r_{\pm})/(r_4-r_{\pm})$, and
\begin{equation*}
 x^2 = \dfrac{(r-r_4)(r_3-r_1)}{(r-r_3)(r_4-r_1)}  , \; \; \; \; \; \; \;  k^2 = \dfrac{(r_3-r_2)(r_4-r_1)}{(r_3-r_1)(r_4-r_2)} ,
 \end{equation*}
 \begin{equation}
B_{+}=\frac{8 M^3r_{+}-4 a^2 M^2-2 a  M \lambda r_{+}}{2 \sqrt{M^2-a^2} \left(r_4-r_{+}\right)} ,
\end{equation}
\begin{equation}
 B_{-}=\frac{-8 M^3 r_{-}+4 a^2 M^2+2 a M \lambda r_{-}}{2 \sqrt{M^2-a^2} \left(r_4-r_{-}\right)}. 
\end{equation}
The functions $F$, $E$ and $\Pi$ appearing in Eq.  (\ref{eq-26}) are the well-known elliptic functions of first, second and third kind, respectively. As the propagation delay for an observer at infinity diverges,  it is usual to compute the time delay w.r.t. a fixed reference point on the orbit, precisely 
\begin{equation}\label{prop_delay}
\arraycolsep=1.6pt\def\arraystretch{1.9}
 \begin{array}{l}
\Delta t_{e x}\left(r_e, \varphi_e\right)=  \left(t_a-t_e\right)-\left(t_a-t_{\text {ref}}\right) \\
=  \left[T_r\left(\infty, \lambda_e\right) \pm T_r\left(r_e, \lambda_e\right) \right]
-\left[T_r\left(\infty, \lambda_{\text {ref }} \right) \pm T_r(r_{\text {ref }}, \lambda_{\text {ref }})\right],
\end{array}   
\end{equation}
where $\lambda_{ref}$ is the angular momentum at the reference point. The ascending node w.r.t. the plane of the sky is usually used as the reference point, as here many post-Newtonian delays are zero. \\
In order to  compute Eq. (\ref{eq24}), we need the coordinates of the emission point $(r_e,\varphi_e)$ in terms of the parameters of the pulsar's orbit as well as its impact parameter $\lambda_e$. As pointed out by \cite{Eva_2019}, there is no general analytical solution to such a problem. For the restricted case of equatorial orbits, one needs to numerically solve \cite{Carleo:2023qxu}
\begin{equation}\label{delta_phi}
    \varphi_a-\varphi_e = \int_{\gamma_r}\dfrac{2Mra-a^2\lambda}{\Delta \sqrt{R(r)}} dr + \int_{\gamma_{r}} \dfrac{\lambda}{\sqrt{R(r)}_{q=0}} dr.
\end{equation}
Notice that $\gamma_r$ includes the points $r_e$ and $r_4$: the first can be written as a function of $\varphi_e$, while $r_4$ is known only after fixing a certain $\lambda$. \\

We will consider a binary system composed of a BH of mass $M$ and a pulsar of mass $m_p$. The relative orbit of the pulsar w.r.t. the black hole (also called `binary orbit') is an ellipse described by
\begin{equation}\label{ellipse}
    R=a_R (1-e_R \cos{u})
\end{equation}
where $a_R$ is  the semi-major axis, $e_R$ the eccentricity and $u$ the eccentric anomaly. We also consider possible relativistic corrections to the orbital motion; therefore,  called $M_{tot}=\bar{M}+m_p$ the total mass, the pulsar orbit is given by $r_p=a_p(1-e_p \cos{u})$ where $a_p= (\bar{M}/M_{tot}) a_R$ while $e_p$ is given\footnote{For the relation between $a_R$ and  binding energy $E$ see Eq. (7.2a) in \cite{Damour1986}.} by \cite{Damour1986}
\begin{equation}\label{eq:ep}
    e_p=e_R\Big[1- \frac{Gm_p(m_p-\bar{M})}{2 M_{tot}a_Rc^2} \Big]
\end{equation} 
and similarly for the black hole eccentricity $e_c$.  Note that the azimuthal position $\varphi_e$ of the pulsar is related to the true anomaly $\phi$ (measured from periastron) by \cite{Eva_2019}
\begin{equation}\label{eq:phi} 
\cos\varphi_{e} = -\sin\textit{i}\sin(\omega+\phi)\,.
\end{equation} 
where $\omega$ is the argument of the periastron and $i$ the inclination of the orbit on the sky plane. In the following, we will also need $\mathbf{r}_p$ and $\mathbf{r}_{c}$, namely the positions vectors w.r.t. the binary center-of-mass of the pulsar and BH, respectively. Moreover, we will call $\mathbf{n}$  the unit vector from the binary center-of-mass to the observer.  

\section{The post-Newtonian formula}

In the weak-field approximation, the propagation time delay is the sum of Roemer delay \cite{Rafikov2005}
\begin{equation}\label{eq:Roem}
    \Delta_R= \frac{-\mathbf{n} \cdot \mathbf{r}_p}{c} = \frac{r_p}{c} \sin{i} \sin{(w+\phi)}
\end{equation}
(note the dependence on the \textit{pulsar}'s orbit radius) and the 1PN  Shapiro delay \cite{Lai2005}
\begin{equation}
\Delta_{S}=  \frac{2GM}{c^3} \ln{\frac{1+e_R \cos{\phi}}{1-\sin{i}\sin{\psi}}}.
\end{equation}
If lensing effect is considered, the latter becomes
\begin{equation}
    \Delta_{S}^{(lens)}= - \frac{2GM}{c^3} \ln{\left[\frac{\sqrt{R_{||}^2+R{\pm}^2}-R_{||}}{a_R(1-e_R^2)} \right]}
\end{equation}
where $R_{||}=R\sin{i}\sin{\psi}$ is the binary separation along the line of sight and $R_{\pm}$ are the image positions in the sky. The lensing correction to the leading term in the logarithm argument goes as $\sim (G \bar{M})/(c^2 a_R)$ \cite{Kramer_2021}. In addition to lensing, there are other delays which are sometimes added in order to capture other GR effects on the propagation of the photons, such as the geometric delay $\Delta_{geo}$ \cite{Lai2005} and the frame-dragging delay $\Delta_{FD}$ \cite{Wex1999}. In particular, the first one is given by \cite{Rafikov2005}
\begin{equation}
    \Delta_{geo}= \frac{2GM}{c^3} \left[\frac{\Delta b_{\pm}}{R_E} \right]^2
\end{equation}
where $R_E=\sqrt{4GMR_{CS}|\sin{i}|}/c$  is the Einstein radius ($R_{CS}$ being the value of $R$ at the superior conjunction) and $\Delta b_{\pm} = (1/2) (\pm\sqrt{b_0^2+4R_E^2}-b_0)$. 
All of these effects are naturally taken into account in the full GR treatment.  Notice that the formulas above use the usual post-Newtonian harmonic coordinates system, which is different from Kerr coordinates \cite{Bilel}. In \cite{Eva_2019} it was shown that, near superior conjunction, the addition of lensing and geometric effects gives more accurate results (where the distance between the weak-field formulas and the full GR ones) than a 2PN Shapiro delay formula. Furthermore, as we checked, the frame-dragging delay is negligible at the distances we are interested in, namely systems with binary separations equal or greater than\footnote{This is the value for the Double Pulsar, for which we assumed $M=1.3$ $M_\odot$. The semi-major axis of  PSR J$0514-4002$E is $\simeq 10^8$ M  and specifically corresponds to $10^{12}$ cm. }  $\sim 10^6$ $M$, for which the Roemer and Shapiro delays largely dominate. For this reason, in the numerical comparison, we also adopt the simpler prescription (valid for zero spin) $r_{PN} \rightarrow r - M$, where $r_{PN}$ is the radial coordinate in the post-Newtonian approach and $r=R$ the Kerr radial coordinate, in a similar way to what was done in \cite{Eva_2019} with the difference that here $r$ and $r_{PN}$ are not computed w.r.t. to the center-of-mass but w.r.t. the companion.  \\
A final remark is about the so-called retardation effect. As pointed out in \cite{Rafikov2005}, the orbital motion of the companion during the crossing of the binary system affects the propagation delay, and hence both $\Delta_S^{(lens)}$ and $\Delta_{geo}$ should be corrected. This effect results in the shift of the delay curves by the amount  $\sim a_{c}/c$, where $a_{c}=(m_{p}/M_{tot})  a_R$ is the semi-major orbit of the companion object, and has been observed only in the Double Pulsar system \cite{Kramer_2021}, thanks to high precision data, but it is in general not considered in the literature, therefore we will neglect it. Note that it is not even included in our full GR calculation, as only the use of a `boosted' Kerr/Schwarzschild metric \cite{Kopeikin_1999} could capture it. However, this is beyond the scope of our work, and it is not essential for the purposes of the estimates we want to find. 

\section{The full GR algorithm}

Both the propagation delay (\ref{prop_delay}) and the angle variation (\ref{delta_phi}) can be computed using elliptic integrals. In order to evaluate the delay, for each point on the pulsar's orbit, we need to know the impact parameter $\lambda_e$, which is given by the numerical solution of Eq.(\ref{delta_phi}). However, finding this solution is very computationally intensive, therefore we started from an array of values for $\lambda$ and then for each value, we evaluated the four roots\footnote{In our numerical algorithms, we will just consider the case of four real roots, avoiding complex numbers. When necessary, we used interpolation to extract physical quantities.} of the potential $R(r)$, in order to obtain an array of values for $\Delta \varphi=\varphi_a-\varphi_e=-\varphi_e$ through Eq. (\ref{delta_phi}). Since $\mathbf{R}=\mathbf{r}_p-\mathbf{r}_{c}$,  an estimate of the range of $\lambda$ is given by the geometrical formula \cite{Rafikov2005}
\begin{equation}\label{eq19}
    \lambda^2 \sim b_0^2 = (\mathbf{n}\times R \times \mathbf{n})^2 =\left(R \cos{\psi} \right)^2 + \left(R\sin{\psi}\cos{i} \right)^2
\end{equation}
where $\psi=w+\phi$. This expression neglects both the retardation effect and the lensing effect (see \cite{Rafikov2005} for details), but this is enough to get a rough estimate of $\lambda$ once the values of $a_R$ and $e_R$ have been assigned. In fact, a shortcut to avoid the numerical resolution of Eq. (\ref{delta_phi}) could be the direct use of Eq. (\ref{eq19}). However, we found that even at large binary separations, the small difference between $b_0$ and $\lambda$ noticeably affects the time delay. Even if  $b_0/\lambda \rightarrow 1$ for increasing values of $a_R$, the absolute difference increases causing a changing in the emission position and hence in the computation of the propagation time delay. In other words, it is not so accurate for our purposes and a complete full GR treatment is the best option. To give an example, in the simpler case of $m_p=0$, $a_R=10^6$ $M$ and $M=10 M_{\odot}$ the use of $b_0$ would lead to a maximum absolute difference (at superior conjunction) between the two approaches of $|\Delta t_{ex}-\Delta_R - \Delta_S^{(lens)}-\Delta_{geo}-\Delta_R^{(BH)}|\simeq 10^{-7}$s instead of $\simeq 3 \times 10^{-8}$ s found using the GR couple of values ($\lambda,\varphi_e$). Notice that we need to subtract the Roemer delay of the companion ($\mathbf{r}_{c}$ instead of $\mathbf{r}_{p}$ in Eq.(\ref{eq:Roem})), namely $\Delta_R^{(BH)}$, as the latter, in general\footnote{Even if we consider the pulsar as a test particle for  the propagation delay, its gravitational influence is not neglected for the orbital motion. This approximation of using different approaches for the two cases is common in the literature.}, is not static and the post-Newtonian Roemer delay is usually computed from the center-of-mass, while $\Delta_{ex}$ 'includes' a Roemer delay computed from the center of the companion.   \\ 

In more detail, the  algorithm for the GR approach basically works with two modules\footnote{We used \textsc{Mathematica 12.2}.}.  With the values of $a_R$, $e_R$, $m_p$, $i$ and $w$ (we will assume $a=0$) as inputs, and after choosing an appropriate range $\{\lambda_{min},\lambda_{max}\}$ for $\lambda$ given by the plotted curve $b_0=b_0(\varphi_e)$, the first module returns a discrete evaluation of the  couples $(\lambda,\varphi_e)$  both for direct and indirect trajectories. These couples are used in the second module to compute the exact (GR) propagation time delay (\ref{prop_delay}). We used the ascending node as reference point, which, for an edge-on orbit corresponds to  $\varphi_e=\pi/2$.  Since it belongs to a direct path, we have 
\begin{equation}
T_r\left(\infty, \lambda_{{ref }} \right) \pm T_r(r_{{ref }}, \lambda_{{ref}})  \rightarrow  T_r\left(\infty, \lambda_{{\bar{\varphi}_e}} \right) - T_r(r_{{\bar{\varphi}_e}}, \lambda_{{\bar{\varphi}_e}}) 
\end{equation}
where $\bar{\varphi}_e$ is the closest angle to $\pi/2$ in our discrete list (with a direct trajectory) and $\lambda_{\bar{\varphi}_e}$ and $r_{\bar{\varphi}_e}$ are the corresponding values of impact parameter and radial coordinate.   We also  assume $w=-\pi/2$ which simplifies Eq. (\ref{eq:phi});  the choice of the periastron position does not affect our  order-of-magnitude estimates. As for $\Delta_{ex}$, for each of the delays  $\Delta_R$, $\Delta_S^{(lens)}$ and $\Delta_{geo}$ we subtract the corresponding delay in the reference point, e.g. $\Delta_{R}(\lambda,\varphi_e) \rightarrow \Delta_{R}(\lambda,\varphi_e) - \Delta_{R}(\lambda_{\bar{\varphi}_e},\bar{\varphi}_e)$. Note that since we only measure differences in times, the reference point can be any point on the orbit, as long as it is the same for all delays. With the latter, we finally compute the absolute difference 
\begin{equation}\label{DIF}
\Delta_{DIF}=|\Delta t_{ex}-\Delta_R - \Delta_S^{(lens)}-\Delta_{geo}-\Delta_R^{(BH)}| 
\end{equation}
which is evaluated at the point closest to the superior conjunction, where we expect the discrepancy to be maximum, especially for highly-inclined orbits. We will sometimes also compute the delay difference far from this portion of the orbit, as for example when the propagation is direct. In this case, we will use
\begin{equation}\label{DIF2}
\Delta_{DIF}=|\Delta t_{ex}-\Delta_R - \Delta_S^{}-\Delta_{2PN}-\Delta_R^{(BH)}|
\end{equation}
where $\Delta_{2PN}$ is the second order Shapiro delay (see Eq. (32) in  \cite{Eva_2019}). \\
We conclude this section with a remark on the case of generally inclined orbits, i.e. $i\not=\pi/2$. If $e_R=0$, the couples $(\lambda,\varphi_e)$ found with the first module does not depend\footnote{Obviously, the  position of the pulsar along its orbit, which is given by the true anomaly $\phi$, does depend on $i$.} on the inclination angle $i$ ($r_e$ is constant), even if, depending on $i$, the range of $\varphi_e$ is limited as $\arccos{(\sin{i})} \leq \varphi_e \leq \arccos{(-\sin{i})}$, meaning that some observation angles $\Delta \varphi$ are not possible.  Since we are neglecting the spin to calculate our estimates, we practically find ourselves in a situation of spherical symmetry. Therefore, to compare the two approaches and find the maximum difference when $i\not=\pi/2$, we decided to proceed in the following way: we  fix $\phi=\pi$ (superior conjunction), find the corresponding couple ($\lambda$,$\varphi_e$) and with this we compute the $\Delta_{ex}$. In other words, thanks to the spherical symmetry, when $i\not=\pi/2$ the exact delay at superior conjunction is equivalent to the exact delay of photons with an emission point at $\varphi_e=\pi/2+i$ on an edge-on orbit. This analogy clearly holds only for circular orbits.  

\section{The relativistic case} \label{sec:rel}

While in the previous works \cite{Eva_2019,Bilel,Carleo:2023qxu}, the pulsar simply follows a Keplerian orbit, here we generalize by allowing  the presence of relativistic corrections to the orbital motion. We  encode the latter with the usual dimensionless parameter $\epsilon=(GM_{tot})/(a_R c^2)$ and use the quasi-Newtonian scheme of Damour\&Deruelle \cite{Damour_I} valid at 1PN. In the relativistic case, the eccentricities of the three orbits (pulsar, black hole, binary)  are not the same (see Eq.(\ref{eq:ep})); however, we still have $r_c+r_p=R$, since $\bar{M} e_p + m_p e_c=M_{tot} e_R$. although there are different ways to describe the relativistic orbit (which is not a closed ellipse) depending on the orbital angle you want to use, the most suitable one is provided by Eq. (\ref{ellipse}), with the important difference that the relation between the true anomaly $\phi$ and the eccentric anomaly $u$ is given by \cite{Damour_I}

\begin{equation}\label{eq25}
    \phi = 2 q \arctan \left( \frac{1 + e_{\theta}}{1 - e_{\theta}}\right)^{1/2} \tan{\frac{u}{2}}
\end{equation}
where 
\begin{equation}
    e_{\theta} = e_R \left(1 + \frac{\epsilon m_p \bar{M}}{2 M_{tot}^2}\right)
\end{equation}
and $ q = 1 + 3 \epsilon/(1-e_R^2)$. To obtain the latter relation, we started from Eq. (4.14) of \cite{Damour_I}, wrote the binary angular momentum by unit reduced mass, $J$, as $J=\sqrt{GM_{tot}a_R(1-e_R^2)}$ and Taylor expanded in $\epsilon$.  With this description,  the periastron advance is only  codified in the constant $q$, which becomes exactly 1 in the non-relativistic case; we  are still allowed to use the constant value $w=-\pi/2$  in Eqs. (\ref{eq:phi}) and (\ref{eq:Roem}) and in the definition of $\psi$,  with the caveat that now it  corresponds to the periastron angle at a specific (proper or coordinate) reference time. \\
The numerical algorithm is similar to the non-relativistic case. Given a suitable range for $\lambda$, a first module finds the roots $r_i$  and then solves two numerical equations (one for direct and one for indirect trajectory) in the variable $-\pi \leq u \leq \pi$. With this,   the angle $\phi$ (see Eq. (\ref{eq25})) and hence $\varphi_e$ (see Eq. (\ref{eq:varphi})) are computed. In a second module, the couples $(\lambda,\varphi_e)$ are used to find the exact relativistic delay $\Delta_{ex}^{(rel)}$. In this case, we are not interested in the comparison with the weak-field approximations, but we want to test the ability of our algorithm to capture the relativistic corrections to the orbital motion, therefore we compute the quantity
\begin{equation}\label{rel}
    \Delta_{relativistic}= \Delta_{ex}^{(rel)} - \Delta_{ex}
\end{equation}
which is identically zero if $\epsilon=0$. With this, we can quantify the importance of including the relativistic corrections (1PN) to the orbital motion. Since these corrections are more pronounced in eccentric systems, we will assume $e_R\not=0$ to evaluate Eq. (\ref{rel}).

\section{Results}
In this section, we report the results found with our algorithm to compare the maximum difference (at superior conjunction) between the full GR propagation delay and the sum of the weak-field delays. We first give some estimates for the simple case of $m_p=0$ (as assumed in previous works). We found that for a pulsar orbiting around a black hole of mass $\bar{M}=10 M_{\odot}$ on a circular, edge-on orbit of radius $a_R=10^6 M$, the maximum (at superior conjunction) absolute difference  is $\Delta_{DIF}\sim  10^{-7}$ s, which is superior to the predicted precision of SKA. This difference decays quickly as $a_R$ increases: when $a_R=10^{7} M$ it is $\Delta_{DIF}\sim \times 10^{-8}$ s, while $\Delta_{DIF}\sim \times 10^{-9}$ s when $a_R=10^{8} M$. This means that if the  stellar BH has a  mass  of $10 M_{\odot}$, the weak-field approximations will be practically indistinguishable from the  full GR formula if  $a_R > 10^{7} M$. Incidentally, we point out that using $b_0$ instead of $\lambda$ would overestimate $\Delta_{DIF}$ by a factor of about 10 (when $a_R=10^{6} M$ and it increases with $a_R$), demonstrating the fact that having maximum precision on lambda is essential. \\

\begin{figure}
\includegraphics[width=\columnwidth]{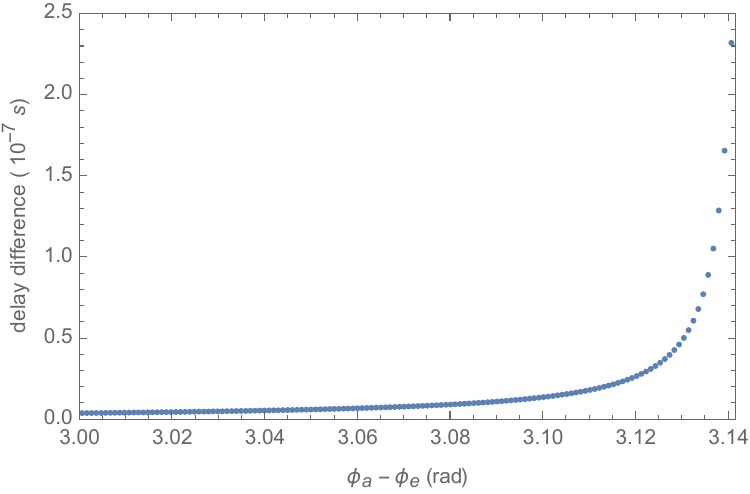}
   \caption{Delay difference $\Delta_{DIF}$ as a function of the observation angle $\Delta \varphi = \varphi_a - \varphi_e \equiv \phi_a -\phi_e$   for the case of a pulsar with mass $m_p=2 M_{\odot}$ orbiting a black hole with mass $\bar{M}=10 M_{\odot}$ on a circular edge-on orbit  ($i=\pi/2$) of radius $a_R=10^6 M \simeq    1.5 \times 10^{12}$ cm  and without relativistic corrections.}
   \label{fig:1}
\end{figure}

\begin{figure}
\includegraphics[width=\columnwidth]{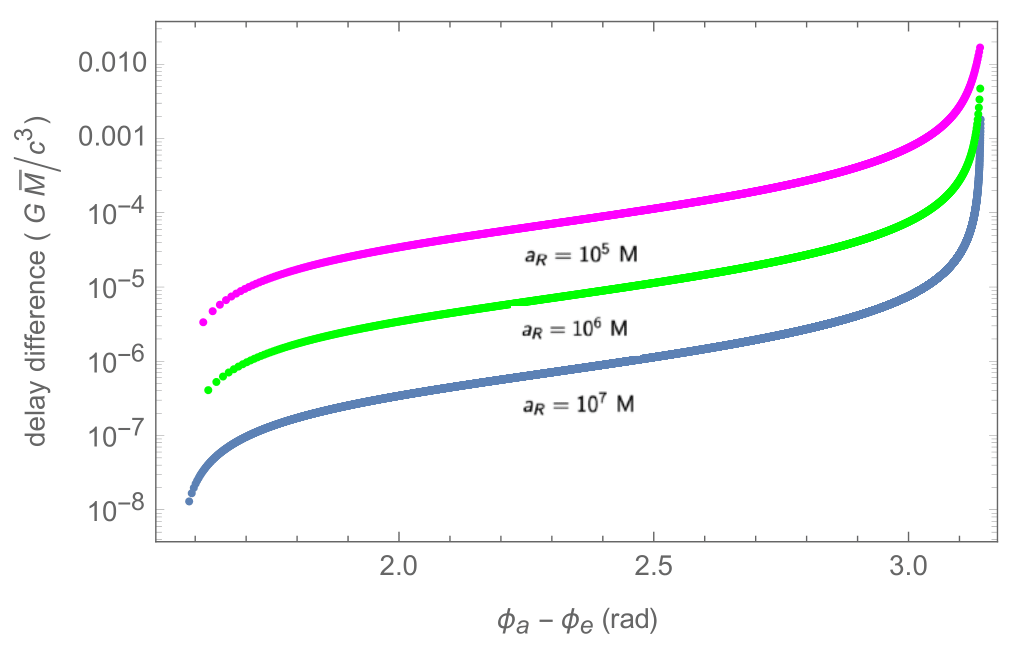}
   \caption{Delay difference $\Delta_{DIF}$ as a function of the observation angle $\Delta \varphi = \varphi_a - \varphi_e \equiv \phi_a -\phi_e $ in units of $G\bar{M}/c^3$ for three different value of the semi-major axis of the relative orbit $a_R$.  }
   \label{fig:2}
\end{figure}

\begin{figure}
\includegraphics[width=\columnwidth]{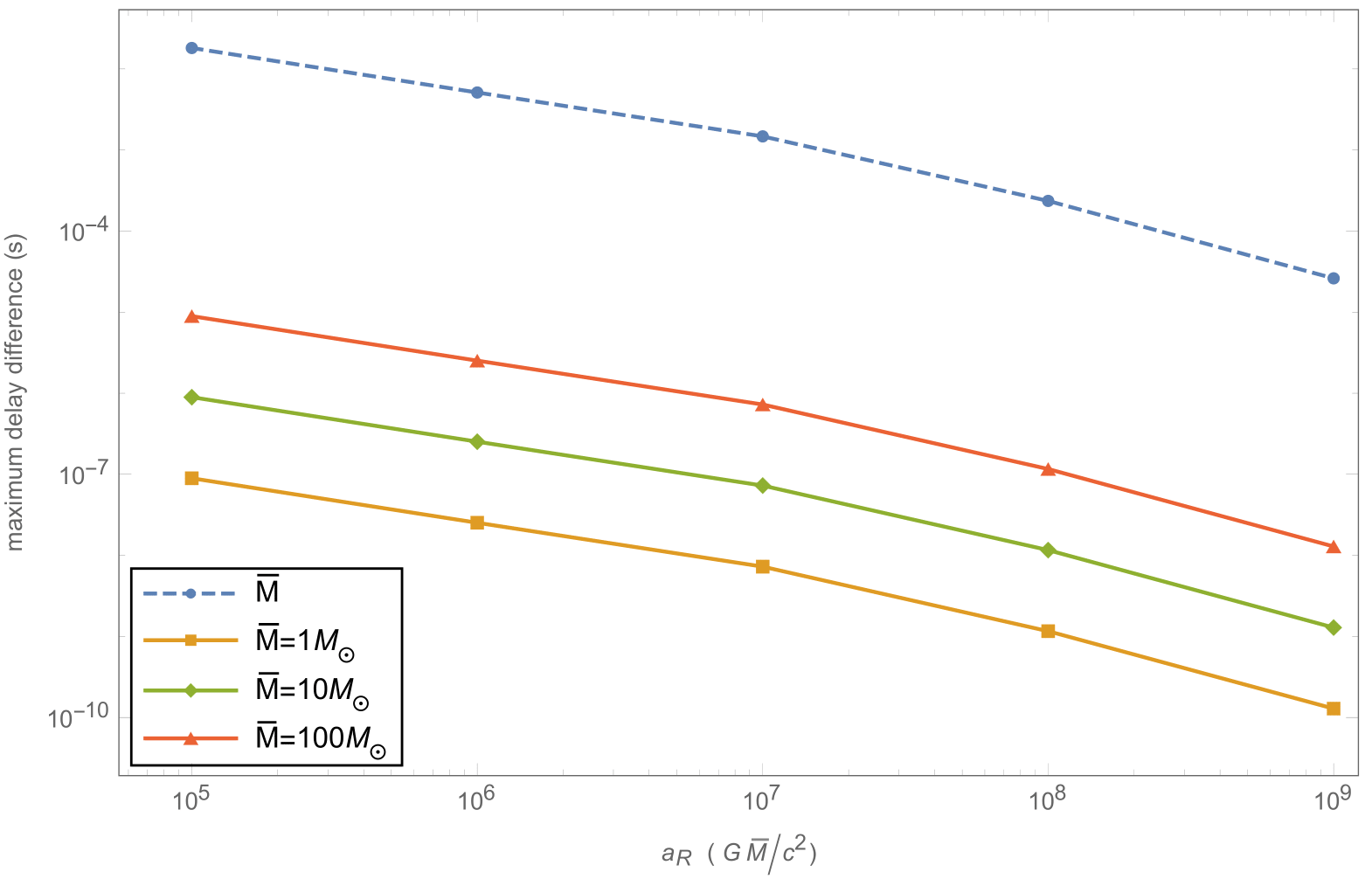}
   \caption{Log-log plot of the maximum delay difference  as a function of the binary semi-major axis $a_R$ for different values of the black hole mass $\bar{M}$. Interpolation was used on a discrete set of couples ($\Delta_{DIF}$, $a_R$).   Notice that the blue dashed line is for a generic mass, therefore the value on the y-axis  must be multiplied by $G \bar{M} c^{-3} \simeq 4.4 \times 10^{-5}$ ($\bar{M}$/$M_{\odot}$).}
   \label{fig:complete}
\end{figure}

\begin{figure}
\centering
    \includegraphics[width=0.48\textwidth]{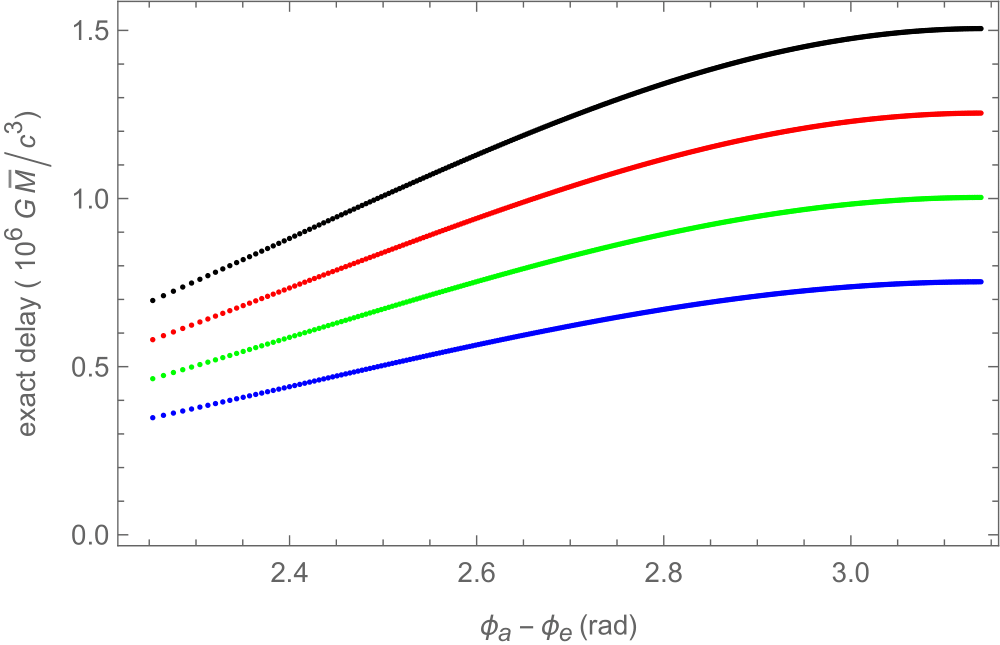}
    \caption{Exact propagation delay $\Delta t_{ex}-\Delta_R^{(BH)}$ for different values of $m_p$: 0 (black), 0.2$\bar{M}$ (red), 0.5$\bar{M}$ and $\bar{M}$ (blue). The other parameters are: $a_R=10^6$ M, $e_R=0.5$, $i=\pi/2$, $w_0=-\pi/2$. The reference point is at $\phi_a-\phi_e=\pi/2$.}
\label{fig:c312}
\end{figure}

In Fig. (\ref{fig:1}), we show the delay difference $\Delta_{DIF}$ for the case of a pulsar with mass $m_p=2 M_{\odot}$ orbiting a black hole with mass $\bar{M}=10 M_{\odot}$ on a circular edge-on orbit  ($i=\pi/2$) of radius $a_R=10^6 M \simeq    1.5 \times 10^{12}$ cm  and without relativistic corrections, i.e. $\epsilon=0$. For simplicity we assumed $w=-\pi/2$ and, as mentioned above, this choice does not affect the order of magnitude of our results. By construction, $\Delta_{DIF} \rightarrow 0$ when $\Delta \varphi \rightarrow \frac{\pi}{2}$; for this reason we plotted only a small region around the superior conjunction. As we can see, $\Delta_{DIF}$ increases rapidly as the conjunction approaches, reaching an absolute value of $\sim10^{-7}$ s, mainly due to the discrepancy in the lensing path which is not so accurately computed in the weak-field approximations. For completeness, we also estimated $\Delta_{DIF}$ at inferior conjunction, where GR effects on the propagation path are deeply suppressed, and we found $\Delta_{DIF}\sim10^{-10}$ s which is noticeably smaller (in this case, we used Eq. (\ref{DIF2}) instead of Eq. (\ref{DIF})). \\
A more general plot is given in Fig. (\ref{fig:2}), where we show the results in units of $G\bar{M}/c^3$, i.e. without assuming a priori a mass for the companion. Three different semi-major axis values are displayed, namely  $a_R=10^5 M$, $a_R=10^6 M$ and $a_R=10^7 M$.  A strong dependence on the mass of the companion appears. For example, if $\bar{M}=10 M_{\odot}$ and $\Delta \varphi = 3$ (near superior conjunction), then $\Delta_{DIF} \sim 10^{-10}$ s if $a_R\simeq10^{13}$ cm, $\Delta_{DIF} \sim 10^{-9}$ s if $a_R\simeq10^{12}$ cm and $\Delta_{DIF} \sim 10^{-8}$ s if $a_R\simeq10^{11}$ cm, meaning that  $\Delta_{DIF}$ increases by one order of magnitude if $a_R$ decreases  by a similar factor. On the contrary, with $a_R$ fixed, $\Delta_{DIF}$ grows linearly with $\bar{M}$. This is all more clear in Fig. (\ref{fig:complete}), where we show the maximum delay difference in seconds as a function of $a_R$ for different choices of $\bar{M}$. Since  we are only interested in the order of magnitude, we used interpolation on a discrete set of couples ($\Delta_{DIF}$, $a_R$). Note that the blue dashed line is for a generic mass, therefore the value on the y-axis  must be multiplied by $G \bar{M} c^{-3}$ in cgs units. From this log-log plot we roughly extract a trend for the maximum difference of the type $\Delta_{DIF} \sim k_{\bar{M}} a_R^{-4/3}$ where $k_{\bar{M}}$ is a constant depending only on the mass of the black hole $\bar{M}$. Indeed, the pulsar's mass $m_p$ is irrelevant for these estimates: different values of $m_p$ certainly imply different orbits for the two objects (and therefore a different propagation delay, see Fig. (\ref{fig:c312})), however $\Delta_{DIF}$, which is basically a difference between propagation delays,  remains practically constant, depending only on $a_R$ and $\bar{M}$, i.e. on the distance to the companion and on the mass of the latter. Finally, we note that, if the mass $\bar{M}$ is sufficiently large ($100 M_{\odot}$), then $\Delta_{DIF}$  is significant ($\sim10^{-7}$ s) even for large binaries  ($a_R\sim10^{16}$ cm) and that, with the very high value  $\bar{M}=10^5 M_{\odot}$, $\Delta_{DIF}$ increases up to    $\sim 10^{-4}$ s.    \\

\begin{figure}
    \centering
    \includegraphics[width=0.48\textwidth]{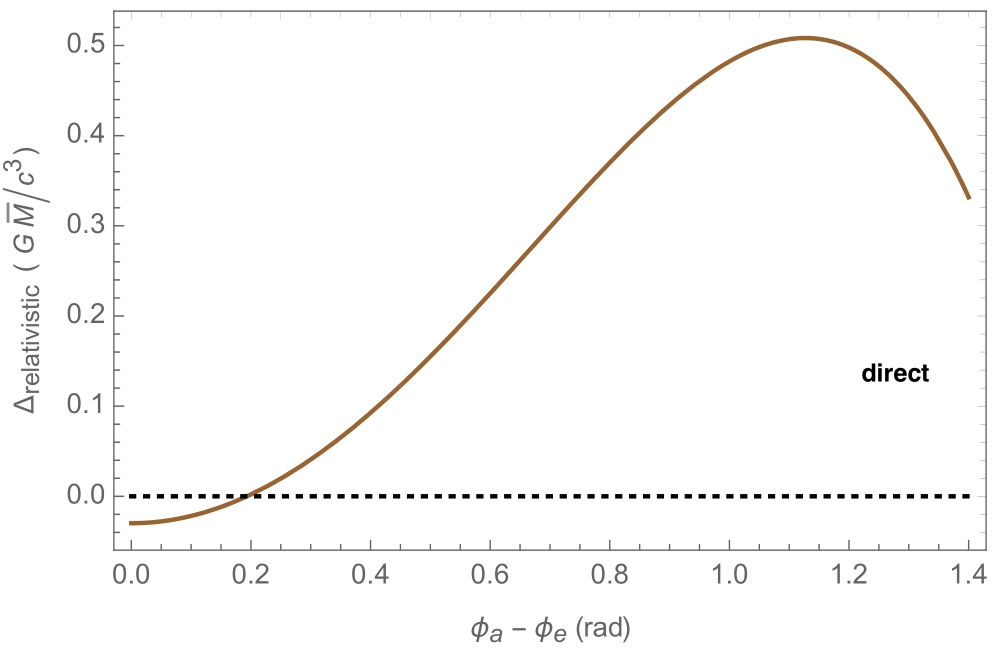}
    \includegraphics[width=0.48\textwidth]{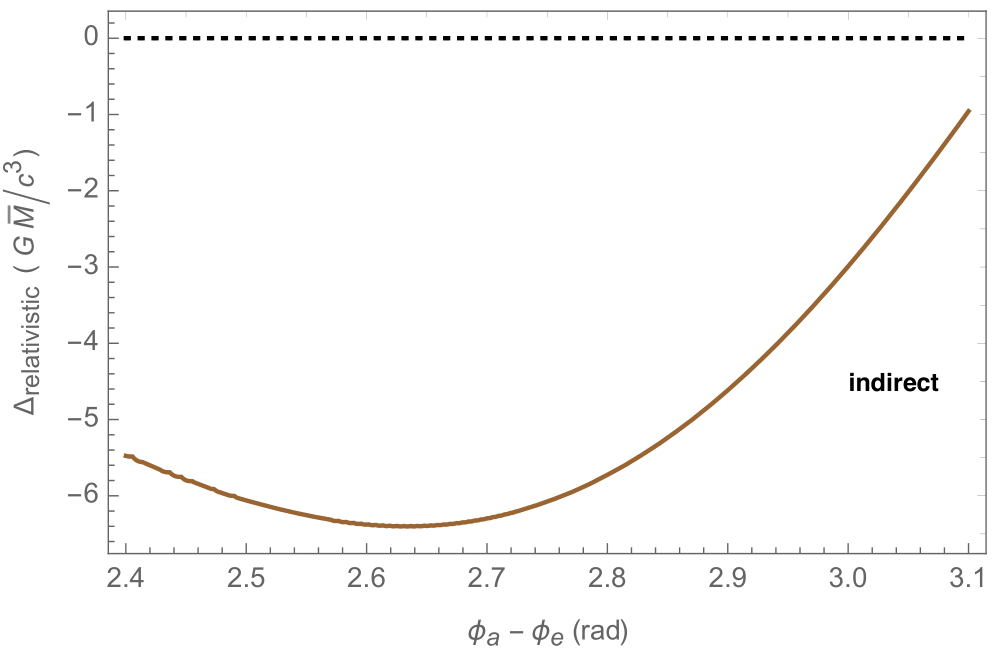}
    \caption{Difference between the exact delay with and without the relativistic corrections (1PN) to the orbital motion, (see Eq. (\ref{rel})) as a function of the observation angle. For the binary system we assumed $a_R=10^5 M$, $e_R=0.5$, $w=-\pi/2$, $i=\pi/2$ and $m_p=0.2 \bar{M}$, where $\bar{M}$ is the mass of the black hole. (Top) Direct propagation. (Bottom) Indirect propagation.}
    \label{fig:4}
\end{figure}

For the relativistic case (Sec. \ref{sec:rel}), i.e. considering also relativistic corrections to the orbital motion, we assume as a toy model a binary system with $a_R=10^5 M$, $e_R=0.5$, $w=-\pi/2$, $i=\pi/2$ and $m_p=0.2 M$. We chose a smaller semi-major axis since we want to emphasize the relativistic contribution. In  Fig. (\ref{fig:4}) we plotted the difference between the exact delay with and without the relativistic corrections (1PN) to the orbital motion, namely Eq. (\ref{rel}). We found that our algorithm is sensitive to relativistic corrections, and these are evident even just considering a single orbit. Depending on the mass of the black hole $\bar{M}$, the relativistic correction can be very important: for a black hole with $\bar{M}=10 M_{\odot}$, the relativistic contribution goes up to $\simeq - 6.5 G \bar{M} c^{-2} \simeq 3.2 \times 10^{-4}$ s. Even if this holds for $a_R=10^5 M \simeq 0.01$ AU, we conclude that neglecting the relativistic corrections can induce very high errors in the time delay. Moreover, since the periastron advance is a secular effect, we expect that this discrepancy increases with the observation time. Therefore,  we recommend always taking into consideration the relativistic contributions, whose importance could be very relevant for future radio telescopes.  \\

Although we expect maximum discrepancies for edge-on orbits, we have also explored cases with a different inclination, in order to decide the importance of using a full GR approach for inclined orbits.  We report the results in Table \ref{tab:1}, where we assumed $a_R=10^6 M$ and three different masses for the black hole, namely $\bar{M}=5 M_{\odot}$, $\bar{M}=10 M_{\odot}$ or  $\bar{M}=100 M_{\odot}$. We  deduce that if $\bar{M}=5 M_{\odot}$  it will be impossible to distinguish the two approaches, except for edge-on orbits. If $\bar{M}=10 M_{\odot}$, instead, orbits with $75° < i < 90°$ could require a full GR computation. Finally, if $\bar{M}=100 M_{\odot}$, then the discrepancy could be relevant already at low inclinations.  Results for different values of $a_R$ scale as discussed above.  

\begin{table*}
  \centering
\begin{tabular}{|p{2.5cm}||p{2cm}|p{2.5cm}|p{2.5cm}|p{2.5cm}|p{2.5cm}|}
 \hline
 \multicolumn{6}{|c|}{Maximum difference delay} \\
 \hline
 Inclination (°) & $\varphi_e $ (rad)  & $\Delta_{DIF}$ $(G\bar{M}c^{-3})$ & $\Delta_{DIF}$ (s) & $\Delta_{DIF}$ (s) & $\Delta_{DIF}$ (s)  \\
 \hline
 15 & $\simeq 1.83$  & $2 \times 10^{-6} $ &  $5 \times 10^{-11} $ &   $1 \times 10^{-10} $  & $1 \times 10^{-9} $ \\
 30 &   $\simeq 2.09$  & $4 \times 10^{-6} $ &$1 \times 10^{-10} $ &  $2 \times 10^{-10} $  & $2 \times 10^{-9} $   \\
 45 & $\simeq 2.35$ & $8 \times 10^{-6} $&  $2 \times 10^{-10} $ &  $4 \times 10^{-10} $  &  $4 \times 10^{-9} $\\
 60 & $\simeq 2.61$ & $1 \times 10^{-5} $&  $4 \times 10^{-10} $ &  $8 \times 10^{-10} $  & $8\times 10^{-9} $ \\
 75 &  $\simeq 2.88$ & $4 \times 10^{-5} $  & $9 \times 10^{-10} $ &  $2 \times 10^{-9} $ &  $2 \times 10^{-8} $ \\
 90 & $\simeq 3.14$  & $5 \times 10^{-3} $ & $1 \times 10^{-7} $&  $2 \times 10^{-7} $ &  $2 \times 10^{-6} $ \\
 \hline
\end{tabular}\caption{Maximum delay difference for different combinations of inclination $i$ and masses $\bar{M}$ of the companion. The third column is in geometric units, while the last three columns are in seconds and for a black hole of mass ${5 M_{\odot}}$, ${10 M_{\odot}}$ or ${100 M_{\odot}}$ respectively. The emission angle $\varphi_e$ is simply given by $\varphi_e=\frac{\pi}{2} + i$. The other orbital parameters are $a_R=10^6 M$, $e_R=0$ and $w_0=-\pi/2$. The results do not depend on the pulsar mass.  }
  \label{tab:1} 
\end{table*}

\section{Summary and outlook}

In the last few years, the timing precision of some pulsar experiments has increased from $\sim10$ $\mu$s to  $\sim100$ ns \cite{Wang_2024}, with good reasons to expect residuals below $\sim50$ ns in the coming years \cite{Hu_2022}, even near superior conjunction, where accuracy is highly compromised due to eclipses. Even if software strategies play an important role in precision improvement, the real leap in quality will come with the entry into operation of SKA, which will be able to achieve an accuracy of $\sim10$ ns \cite{Smits_2011,Liu_2011}. Two important goals of SKA are: (1) to detect and perform the timing of pulsars in the galactic center, which is a discovery that has been awaited for many years and which has numerous scientific implications; (2)  to find and time with a great precision a pulsar in orbit around a \textit{stellar} black hole \cite{Desvignes_2016}, which we expect to find in particular in globular clusters. The strong-field regime in the galactic center region has led us to question whether the usual post-Newtonian formulas used for the timing are still valid. A series of papers \cite{Eva_2019,Bilel,Carleo:2023qxu,DellaMonica:2023ydm} have confirmed that the timing of a pulsar in the galactic center requires a full GR approach to calculate the propagation delay, especially for edge-on orbits. For Sgr A*, the maximum difference (superior conjunction) between the exact (analytical) formula and the sum of the single post-Newtonian delays is of $\sim1$ s for a pulsar orbiting at distance $r=10^2-10^3$ M. A common feature of these works is the assumption of a pulsar orbiting a SMBH, therefore allowing the use of the one-body approximation, since the neutron star is considered as a test particle in the space-time of the black hole. This circumstance, however, in addition to being unique (there is only one supermassive black hole in our galaxy) still does not have an observational counterpart, while the number of pulsars orbiting stellar black holes ($5-100$ $M_{\odot}$) should be relatively higher, especially in globular clusters \cite{Kremer_2018}, where we also expect intermediate-mass black holes ($10^2-10^5$ $M_{\odot}$) \cite{Fujii_2024}, for which the accuracy of the post-Newtonian approximations had not yet been investigated. For these reasons, we have undertaken an analysis  on the discrepancy in the propagation delay (the most substantial part of TOAs) between the usual post-Newtonian formula and a full GR approach, having in mind this close-to-discovery class of systems, i.e. binary pulsars with a stellar or intermediate black hole as a companion. This generalization has made us consider the importance of the orbital motion of the black hole companion. We have therefore corrected the procedure found in these previous works and developed an algorithm to compare the two approaches in the case of these new systems. In addition, we have also considered relativistic corrections to the orbital motion, which have been neglected in the aforementioned works for simplicity. \\
In more detail, the algorithm for the GR approach basically works with two modules. Given the main orbital parameters as input, the first module returns a discrete evaluation of the impact parameter and emission angle of the photons, while the  second module computes the exact (GR) propagation time delay using Eq. (\ref{prop_delay}) and  the absolute difference (\ref{DIF}) or (\ref{DIF2}), depending on whether the trajectory is direct or indirect.  We found that for a pulsar orbiting a black hole of mass $\bar{M}=10 M_{\odot}$ on a circular, edge-on orbit of radius $a_R=10^6 M$, the maximum (at superior conjunction) absolute difference is $\Delta_{DIF}\sim  10^{-7}$ s, which is superior to the predicted precision of SKA, causing non-zero residuals. This difference decays quickly as $a_R$ increases: when $a_R=10^{7} M$ it is $\Delta_{DIF}\sim \times 10^{-8}$ s, while $\Delta_{DIF}\sim \times 10^{-9}$ s when $a_R=10^{8} M$. This means that if the stellar BH has a  mass  of $10 M_{\odot}$, the weak-field approximations will be practically indistinguishable from the full GR formula if  $a_R > 10^{7} M$. In general, we found that the difference between the two approaches, $\Delta_{DIF}$, increases by about one order of magnitude if the semi-major axis of the relative orbit, $a_R$, decreases  by a similar factor.  More precisely, we roughly extracted a trend for the maximum difference of the type $\Delta_{DIF} \sim k_{\bar{M}} a_R^{-4/3}$, where $k_{\bar{M}}$ is a constant depending only on the mass of the black hole $\bar{M}$. On the contrary, with $a_R$ fixed, $\Delta_{DIF}$ grows linearly with the companion's mass $\bar{M}$, as clear in Fig. (\ref{fig:complete}), where we show the maximum delay difference (in seconds) as a function of $a_R$ for different choices of $\bar{M}$. About the pulsar's mass $m_p$, we discovered that it is basically irrelevant for our estimates on $\Delta_{DIF}$: different values of $m_p$ certainly imply different orbits for the two objects (and therefore a different propagation delay, see Fig. (\ref{fig:c312})), however $\Delta_{DIF}$, which is  a difference between propagation delays,  remains practically constant, depending only on $a_R$ and $\bar{M}$, i.e. on the distance to the companion and on the mass of the latter.  This is the result of the theoretical framework adopted in this work and in the previous ones, according to which the pulsar is not considered for the geometry of the space-time in which photons travel; this approximation is always used in the literature due to the lack of an exact analytical metric for binary systems in general relativity.  The consequence is  that the turning on of the pulsar's mass in the computation of the orbital motions does not force us to deeply modify the theoretical computation of the propagation delay. The general outline for the comparison with the post-Newtonian prediction remains mostly the same as that of previous works, provided that the parameters of the \textit{relative} orbit are used for the computation of the delay. Speaking of more massive black holes, if the mass $\bar{M}$ is sufficiently large ($100 M_{\odot}$), then $\Delta_{DIF}$  is significant ($\sim10^{-7}$ s) even for large binaries  ($a_R\sim10^{16}$ cm) and, with the very high value  $\bar{M}=10^5 M_{\odot}$, it increases up to    $\sim 10^{-4}$ s. 
For the relativistic case (see Eq. (\ref{rel})) instead, we found that, depending on the mass of the black hole $\bar{M}$, the relativistic correction can be very important: for a black hole with $\bar{M}=10 M_{\odot}$, the relativistic contribution goes up to $\simeq - 6.5 G \bar{M} c^{-2} \simeq 3.2 \times 10^{-4}$ s  if $a_R=10^5 M$. Therefore, we conclude that neglecting the relativistic corrections can induce very high errors in the time delay, already in a single-orbit simulation. Since the periastron advance is a secular effect, we expect that this discrepancy increases with the observation time. Finally, although we expect maximum discrepancies for edge-on orbits, we have also explored cases with a different inclination, in order to decide the importance of using a full GR approach for inclined orbits.  We report the results in Table \ref{tab:1}, where we assumed $a_R=10^6 M$ and three different masses for the black hole, namely $\bar{M}=5 M_{\odot}$, $\bar{M}=10 M_{\odot}$ or  $\bar{M}=100 M_{\odot}$. We  deduce that if $\bar{M}=5 M_{\odot}$  it will be impossible to distinguish the two approaches, except for edge-on orbits. If $\bar{M}=10 M_{\odot}$, instead, orbits with $75° < i < 90°$ could require a full GR computation. Finally, if $\bar{M}=100 M_{\odot}$, then the discrepancy could be relevant even at low inclinations.  \\

There are several ways to extend  the results of this work. An important improvement would be the extension of a fully relativistic approach to systems with objects of comparable masses, such as the case of the Double Pulsar. First of all, to calculate the trajectory of the photons, we assumed the space-time as generated only by the companion object. Although this is a usual feature in the field of pulsar timing, this consideration is substantially wrong in cases where the two objects have comparable masses and only works at the 1PN order, where the pulsar's potential introduces a nearly constant delay along the orbit, which is therefore not measurable.  However, it is clear that the space-time of relativistic binaries cannot be accurately described by a 1PN 2-body metric, which is not able to capture the complex interaction. On the other hand, i.e. considering a fully analytical approach, it is well known that for binary systems, there are no exact solutions in GR. Therefore, for such systems, we still need to find a middle ground between the 1PN metric (which is not sufficient) and a fully relativistic metric (which does not exist). We ultimately expect that the propagation delay should depend on the pulsar mass and not just the mass of the companion. This can be achieved by considering approximations of the binary metric that contain interaction terms. Even if it would not be a fully relativistic approach, we believe that this better description of the \textit{binary} space-time may result in non-negligible discrepancies for future timing experiments on systems like the Double pulsar or PSR J$0514-4002$E. \\  
An easier extension, instead, could be the addition of higher-order (2PN) relativistic corrections (only) to the orbital motions, which could have a stronger impact on the accuracy of the timing model than the post-Newtonian approximations of the space-time, since in this work we noticed that the precision at which the emission point along the orbit is known has a huge impact on the evaluation of the delays. Moreover, since relativistic versions of Roemer and Shapiro delays are known \cite{Damour1986}, it might be fruitful to perform a comparison using them in the post-Newtonian part. However, in this case the need to use  eccentric orbits (relativistic effects manifest themselves more in systems with non-zero eccentricity) could complicate the relation between the orbital parameters ($a_R,e_R,a_p,a_c,e_p,e_c,u$) of the post-Newtonian  space-time and the corresponding quantities in the Schwarzschild space-time, a task that is immediate when $e_R=0$ (simply  by using the rule $r_{PN}\rightarrow r - M$). A more difficult task might be to use a `boosted' metric \cite{Kopeikin_1999} instead of the static one used here and in previous papers; this could be far more involved but it would be the only way to include the so-called retardation effect in a full GR calculation of the propagation delay. Finally, we want to point out that the magnitude of $\Delta_{DIF}$ is not directly comparable with the timing precision, as the latter is the result of a fit on several parameters depending on several conditions (pulse profile, observation time, signal-to-noise ratio, etc..), therefore the impact of the inaccuracy of the current post-Newtonian formulas on the estimation of the physical quantities (e.g. masses) or on the use of the propagation delay to detect deviations from GR may vary from source to source and it would be useful to quantify it with real cases.   

\begin{acknowledgments}
The authors would like to acknowledge the support of the Istituto Nazionale di Astrofisica (INAF). AC thanks Eva Hackmann for several valuable comments. We acknowledge financial support under the National Recovery and Resilience Plan (NRRP), Mission 4, Component 2, Investment 1.1, Call for tender No. 104 published on 2.2.2022 by the Italian Ministry of University and Research (MUR), funded by the European Union - NextGenerationEU - Project Title GUVIRP-Gravity tests in the UltraViolet and InfraRed with Pulsar timing -- CUP C53D23000910006 - Grant Assignment Decree No. 962 adopted on 30/06/2023 by the Italian Ministry of University and Research (MUR).
\end{acknowledgments}

\bibliographystyle{abbrv}
\bibliography{biblio2}

\end{document}